\documentclass[preprint,showpacs,aps]{revtex4} 
\input{epsf}  
\usepackage{graphicx}
\usepackage{epic}
\usepackage{eepic}
\usepackage{amsmath}
\usepackage{eufrak}
\usepackage{color}

\begin{document}  

\title{
Quantifying the connectivity of a network: 
\\
The network correlation function method
} 
\author{Baruch Barzel and Ofer Biham}
  
\affiliation{  
Racah Institute of Physics,   
The Hebrew University,   
Jerusalem 91904,   
Israel}  

\newcommand{\av[1]}
{
\langle #1 \rangle
}

\newcommand{\Ns}
{
\langle N^2 \rangle
}

\newcommand{\mat[3]}
{
{#1}_{#2#3}
}

\newcommand{\NH}[1][]
{
\langle #1 \rangle
}

\newcommand{\W}[1]
{
W_{\rm #1}
}

\newcommand{\A}[1]
{
A_{\rm #1}
}

\newcommand{\f}[1]
{
f_{\rm #1}
}

\newcommand{\R}[1]
{
R_{\rm #1}
}

\newcommand{\Svis}
{
a_{\rm H}/W_{\rm H}
}

\newcommand{\Svac}
{
W_{\rm H}/f_{\rm H}
}

\newcommand{\Prob}[1]
{
P(\dots#1\dots)
}

\newcommand{\dotProb}[1]
{
\dot{P}(\dots#1\dots)
}

\newcommand{\F}
{
F_{\rm cor}
}

\begin{abstract}

Networks are useful for describing systems of interacting objects,
where the nodes represent the objects and the edges represent the
interactions between them.
The applications include chemical and metabolic systems,
food webs as well as social networks.
Lately, it was found that many of these networks display
some common topological features, such as high clustering, 
small average path length (small world networks) 
and a power-law degree distribution (scale free networks).
The topological features of a network are commonly related
to the network's functionality.
However, the topology alone does not account for the nature
of the interactions in the network and their strength. 
Here we introduce a method for evaluating the correlations
between pairs of nodes in the network.
These correlations depend both on the topology and on
the functionality of the network.
A network with high connectivity displays strong correlations
between its interacting nodes and thus features small-world 
functionality.
We quantify the correlations between all pairs of nodes in
the network, and express them as matrix elements in the
correlation matrix. 
From this information one can plot the correlation function
for the network and to extract 
the correlation length.
The connectivity of a network is then defined as the ratio between
this correlation length and the average path length of the network.
Using this method we distinguish between a topological small world and
a functional small world, where the latter is 
characterized by long range correlations and high connectivity.
Clearly, networks which share the same topology, 
may have different connectivities,
based on the nature and strength of their interactions.
The method is demonstrated on  
metabolic networks, but can be readily 
generalized to other types of networks. 

\end{abstract}

\pacs{89.75.Hc,89.75.Fb,89.75.Da}

\maketitle

\section{Introduction}
\label{introduction}

A network, or graph, consists of a set of nodes, from which selected
pairs are connected by edges.
Such mathematical constructions
provide a useful description for systems
of interacting objects. 
More specifically, network concepts are used in the analysis of
chemical and metabolic systems as well as food webs
and social networks.
In recent years, there has been much progress in the analysis of
the topology of these networks.
The network topology can be characterized
by features such as the number of nodes,
$J$,
and the 
average degree
$\av[k]$, 
namely the average number of edges 
that are connected to a node.
A more detailed description of the network 
topology is given by 
the degree distribution, 
$P(k)$,
which is the probability that a randomly selected
node has exactly 
$k$
edges. 
Another important topological feature
measures the 
tendency of a network to support the formation of cliques.
A clique is a fully connected set of nodes,
namely each pair of nodes in such a set is connected by an edge.
The tendency of a network to form cliques can be characterized by the
clustering coefficient 
\cite{Watts1998,Wasserman1994,Barrat2000,Newman2000}.
Roughly speaking, when a network has a high clustering 
coefficient it is considered to be highly connected.
A low clustering coefficient implies that the network 
is only loosely connected. 

Networks exhibit a unique metric, in which 
the distance, $d$, between any two nodes 
is given by the minimal number of edges one
has to cross in order to pass from one node to the other.
In some cases, 
the distance can be used as a measure for the connection 
between a pair of nodes.
This is based on the assumption that
two directly reacting nodes 
($d = 1$)
strongly affect each other, 
whereas distant nodes weakly affect
one another.
The average path length in a network,
$\av[d]$,
is obtained by averaging over the distance between 
all pairs of nodes in the network.
The parameters defined above were evaluated for random graphs
and their dependence on $J$ and $\langle k \rangle$
was found
\cite{Erdos1959,Erdos1960,Erdos1961,Chung2001}.
However, the analysis of realistic networks shows that they 
are very different from random graphs
\cite{Barabasi2001}.
In realistic networks it is common to find 
surprisingly low average path lengths,
and relatively high clustering coefficients.
In many cases the degree distribution follows a power law form, 
rather than the Poisson distribution
which is the signature of random networks.
These features were found to appear in
social networks
\cite{Kochen1989,Watts1998,Newman2001a,Newman2001b,Newman2001c,Barabasi2001,Redner1998,Vazquez2001,Newman2000,Barabasi1999,Albert2000,Amaral2000},
the world wide web
\cite{Lawrence1998,Lawrence1999,Albert1999,Broder2000,Adamic2000,Adamic1999},
ecological networks
\cite{Pimm1991,Williams2000,Montoya2002,Camacho2002a},
and metabolic networks
\cite{Wagner2000,Fell2000,Jeong2000}.

While the topological properties of realistic networks have been elucidated,
the implications on the functionality of these
networks are not fully understood.
The small average path length 
and the high clustering of many realistic networks, 
render them as small world networks.
At first glance,
the small world characteristics imply
that realistic networks function as highly connected systems.
Indeed, one expects that if the distance between two nodes is small,
the correlation between them will be strong.
For instance, in the case of a metabolic network, 
the concentrations of interacting
proteins will strongly depend on each other.
A perturbation in the concentration of one protein is likely 
to affect the concentration of the other.
This might lead to the conclusion that small world networks    
are highly susceptible to local perturbations, 
as almost all the nodes are just a short distance away.
The problem with this topological analysis, 
is that it does not relate to the
specific function of a given network 
or to the strength of the interactions between its nodes
\cite{Barabasi2002}.
Consider, for instance, a metabolic network 
and an ecological network sharing the same 
topology.
In what sense can these two networks be regarded as similar networks?
Even if the two have the same topological structure, the nature of their
functional behavior is fundamentally different.
The process of predation may lead to different behavior than the 
process of chemical reaction between proteins.
Even two metabolic networks may function differently
if the interaction strengths in one network are higher than in the other.

In this paper, we present a method for
obtaining the correlation matrix of a given network.
The elements of this matrix provide the magnitudes of the correlations
between pairs of nodes in the network.
In certain cases the matrix 
can be used to characterize some of the global features of the 
network's functionality.
For instance, it can be used to identify domains of high 
correlations versus domains of low correlations.
Another use of the correlation matrix is in
quantifying the connectivity of a network in a way
that accounts both for its topology and for the specific processes 
taking place between its nodes.
This method, 
referred to as the network correlation function (NCF) method,
enables us to determine 
whether a topological small world (TSW) network 
will also be a functional small world (FSW) network.
A network will be regarded as an FSW network if
the correlations between its nodes are typically high, 
and thus the state of one node 
is highly dependent on that of the others. 
Here we apply
the method to metabolic networks with various topologies
and different interaction strengths.
In these networks, each node represents a reactant, and is assigned
a dynamical variable that accounts for the concentration of this reactant.
The time dependence of these concentrations is described by
a set of rate equations.
The equations include terms that describe 
the interaction processes in the given network. 
They account both for the topology and for the functionality of the network.
From the solution of the rate equations 
under steady state conditions
one can extract the correlation 
between each pair of nodes.
In certain cases, networks are found to have a typical correlation length.
If the distance between two nodes is much higher than this length, 
the correlation between them is negligible.
To quantify the connectivity of the network,
one compares the correlation length with the average path length.
In case that the average 
path length is smaller than the typical correlation length,
the network will be considered as an FSW network. 
In this case, local perturbations will have a global 
effect on the network.
The FSW network will thus be regarded as strongly connected.
On the other hand, if the average path length is larger 
than the typical correlation length, the network will
be considered as weakly connected.

The paper is organized as follows.
In Sec. II we present the methodology, and 
demonstrate its applicability to metabolic networks.
In Sec. III we analyze some simple, analytically soluble networks, and
in Sec. IV we present a computational analysis of a set of more 
complex networks, culminating in an example of a scale free network.
The results are summarized and discussed in Sec. V.


\section{The Method}
\label{sec_2}

Below we present the NCF method for 
evaluating
the connectivity of
interaction networks.
For concreteness, we focus on the specific case of metabolic networks.
It is straightforward to generalize the method
to other types of networks.
Consider a metabolic network consisting of 
$J$
different molecular species,
$X_i$, $i=1,\dots,J$.
The generation rate of the
$X_i$ molecules
is
$g_i$ (s$^{-1}$). 
Once a molecule is formed it may undergo degradation at a rate 
$w_i$ (s$^{-1}$).
Certain pairs of molecules,
$X_i$ and $X_j$, 
may react to form a more complex molecule 
$X_k$ 
($X_i+X_j \rightarrow X_k$).
In general, the product molecules $X_k$ 
may be reactive and represented by another node in 
the network. For simplicity, in the analysis below,
we assume that the $X_k$ molecules are not reactive and
thus do not play a further role in the network.
We also limit the discussion to the case in which a
molecular species does not react with itself, namely
reactions of the form 
$X_i+X_i \rightarrow X_k$
are excluded.

The reaction rate between 
the 
$X_i$ and $X_j$ molecules 
is given by the 
{\it reaction rate matrix}
$A$.
Its matrix elements are
$a_{ij}$ (s$^{-1}$),
where
$i,j = 1,2,\dots,J$.
Note that for non-interacting pairs of molecules
$a_{ij} = 0$.
The {\it network topology matrix},
$\mathcal M$,
is also a
$J \times J$ 
dimensional matrix, 
which is defined as follows:
$\mat[M]ij = 1$
if 
$X_i$ and $X_j$ 
react with each other, and
$\mat[M]ij = 0$
otherwise.
Let 
$\mat[D]ij$ 
be the distance between the species 
$X_i$ and $X_j$
in the metric of the network.
The average path length is thus

\begin{equation}
\av[d] = \frac{1}{J(J-1)} \sum_{i,j=1}^{J}{\mat[D]ij}.
\label{eq:average_distance} 
\end{equation}

\noindent
The parameter
$\av[d]$
provides some information as to the connectivity of the network, 
but only in the topological sense.

In order to account for the functionality of the network we
consider the rate equations,
which take the form

\begin{equation}
{ \frac{dn_i}{dt} } = g_i - w_i n_i(t) - \sum_{j=1}^{J}a_{ij} n_i(t)n_j(t),
\label{eq:metabolic_rate}
\end{equation}

\noindent
where 
$n_i (t)$
is the time dependent concentration of the molecule
$X_i$.
The first term on the right hand side of 
Eq. (\ref{eq:metabolic_rate})
accounts for the generation of 
$X_i$
molecules.
The second term accounts for the process of degradation, 
and the third term accounts for reactions between molecules.
The steady state (SS) solution of the rate equations,
$n_i$, 
can be obtained by setting the left hand side of 
Eq. (\ref{eq:metabolic_rate}) 
to zero. 
One obtains

\begin{equation}
n_i = \frac{g_i}{w_i^{\rm eff}},
\label{eq:metabolic_ss}
\end{equation}

\noindent
where 
$w_i^{\rm eff} = w_i + \sum_{j} a_{ij} n_j$
is the effective degradation rate. 
Our goal is to characterize the correlations between the different species 
around the steady state condition.
Roughly speaking, we are asking the following question:
While at steady state, 
to what extent does a small perturbation 
in the concentration of the species 
$X_j$ 
affect the concentration of the species 
$X_i$?  
To this end we define the 
{\it first order correlation matrix} as 

\begin{equation}
\mat[C]ij = \left. { \frac{\partial n_i}{\partial n_j} } \right| _{\rm SS},
\label{eq:2ni_2nj}
\end{equation}

\noindent
which, using 
Eq. (\ref{eq:metabolic_ss}) 
takes the form

\begin{equation}
\mat[C]ij = - { \frac{ a_{ij} g_i}{({w_i^{\rm eff}})^2}}.
\label{eq:Cij}
\end{equation}

\noindent
Note that the elements of the first order correlation 
matrix are non-zero only if the species 
$X_i$ and $X_j$
directly interact with each other. 
Topologically, this means that the matrix element 
$\mat[C]ij$
vanishes unless
$\mat[D]ij = 1$.
Indirect correlations between species that are connected via a third species 
are not accounted for (hence the term first order correlation matrix).
To account for indirect correlations, 
one has to compute the complete correlation matrix

\begin{equation}
\mat[G]ij = \left. { \frac{dn_i}{dn_j}}\right|_{\rm SS}.
\label{eq:dni_dnj}
\end{equation}

\noindent
Clearly, the diagonal terms of this matrix must satisfy

\begin{equation}
{\left. \frac{dn_i}{dn_i} \right|_{\rm SS} } = 1,
\label{eq:Gii}
\end{equation}

\noindent
for 
$i=1,\dots,J$. 
For the off-diagonal terms,
$i \ne j$,
one can write

\begin{equation}
\left. { \frac{dn_i}{dn_j}} \right|_{\rm SS} = 
{\left. { \frac{\partial n_i}{\partial n_j}} \right|_{\rm SS}}
+ \sum_{\substack{k=1\\k \ne j}}^{J}
{{\left. {\frac{\partial n_i}{\partial n_k}} \right|_{\rm SS}}
\left. {\frac{dn_k}{dn_j}} \right|_{\rm SS} }.
\end{equation}

\noindent
In matrix form, these equations become

\begin{eqnarray}
\left\{
\begin{array}{lrr}
\mat[G]ii = 1 & &
\\
\mat[G]ij = \sum_{k=1}^{J}{\mat[C]ik \mat[G]kj} & \,\, & (i \ne j).
\end{array}
\right.
\label{eq:Gij}
\end{eqnarray}

\noindent
Eq. (\ref{eq:Gij}) is a set of 
$J \times J$
coupled linear equations.
Their solution provides the complete correlation matrix, 
$\mat[G]ij$.

Typically,
one expects the correlation between two species to 
decay as a function of the distance,
$\mat[D]ij$,
between them.
The rate of this decay provides the correlation length. 
To obtain the correlation function we identify all pairs of
species $i$ and $j$ that are separated by a distance $d$ from each other.
We then average the magnitude of the correlations, 
$|\mat[G]ij|$,
over all these pairs.
The correlation function vs. distance takes the form

\begin{equation}
\F(d) = 
\frac
{\sum_{i,j=1}^{J}{\left| \mat[G]ij \right| \delta_{d,\mat[D]ij}} }
{\sum_{i,j=1}^{J} \delta_{d,\mat[D]ij}},
\label{eq:g_of_d}
\end{equation}

\noindent
where 
$d$
is an integer.
The function 
$\delta_{x,y} = 1$
if
$x = y$
and zero otherwise.
Note that in the definition of 
$\F(d)$ 
the absolute value of the matrix terms 
$\mat[G]ij$
was used.
This is because certain pairs of species 
$X_i$
and 
$X_j$ 
may be positively correlated, and others may be negatively correlated.
In any case, the focus here is merely on the strength of their mutual 
correlations and not on the sign of these correlations.

To obtain the correlation length, one may fit the
function $\F(d)$ to an exponent of the form
$K \exp (-d/d_0)$.
The distance
$d_0$
is the correlation length.
It approximates the distance within which strong 
correlations between 
different species are maintained.
This distance is determined by the 
dynamical processes and by the characteristic 
rate constants of a specific network. 
It thus accounts not only
for the topology of the system, but also for its functionality.
Finally, we define the {\it connectivity} of a network as

\begin{equation}
\eta = \frac{d_0}{\av[d]}.
\label{eq:eta}
\end{equation}

\noindent
In the limit where 
$d_0$ 
is much greater than the average path length,
most of the nodes are within the correlation length from
one another, and 
the components of the network are highly correlated.
The concentrations of different species are strongly dependent on each other,
and the network is an FSW network.
Correspondingly, one obtains that
$\eta \gg 1$.
In case that 
$d_0$
is much smaller than
the average path length, 
the effect of a perturbation in the concentration of one species 
decays on average before it reaches 
most of the other species. 
Perturbations are thus local,
and the connectivity of the network is said to be low.
While
topologically, such a network might be considered a small world network,
functionally it is a loosely connected network.


\section{Analytically Soluble Networks}

\subsection{Linear Metabolic Network}
\label{example_1}

\begin{figure}
\includegraphics[width=12cm]{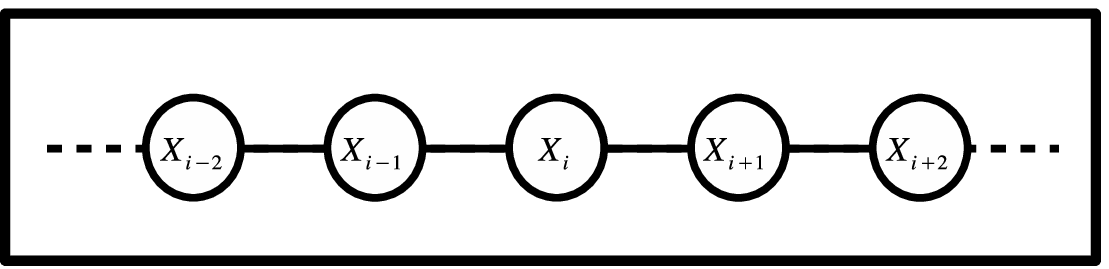}
\caption{
The linear metabolic network.
Each molecular species
$X_i$
reacts with its two nearest neighbors,
$X_{i-1}$ and $X_{i+1}$.
}
\label{fig1_linear_network}
\end{figure}

To demonstrate the NCF method we now refer
to a set of simple examples, which are analytically soluble.
Consider a linear metabolic network of 
$J$ 
species 
($J \gg 1$).
The species 
$X_i$, $i = 1, \dots J$,
reacts with its nearest neighbors, namely
$X_{i-1}$ and $X_{i+1}$.
This network is shown in 
Fig. \ref{fig1_linear_network}.
For simplicity, we take all the reacting species 
to have identical parameters, 
namely 
$g_i = g$
and 
$w_i = w$ 
for 
$i = 1, \dots J$.
Also,
$a_{ij} = a$ 
in case that 
$i = j \pm 1$, 
and 
$a_{ij} = 0$
otherwise.
Taking the limit in which the number of species 
$J$
is very large, 
we can avoid the complexities 
related to the boundaries of the network.
Under these conditions, the steady state 
solution for all the species is the same,
enabling us to omit the index 
$i$ 
from the steady state concentrations
$n_i$.
The reaction rate matrix for this network is 

\begin{equation}
A = 
\left(
\begin{array}{cccccc}
0      & a     & 0      & \dots  &       & 0      \\
a      & 0     & a      &        &       & 0      \\
0      & a     & 0      & a      &       & 0      \\
\vdots &       &        & \ddots &       & \vdots \\
0      & 0     & \dots  & a      & 0     & a      \\
0      & 0     & \dots  &        & a     & 0      \\
\end{array}
\right).
\end{equation}

\noindent
For a linear network, 
the average distance between pairs is 
$\av[d] = (J+1)/3$, 
which for 
$J \gg 1$,
can be approximated by

\begin{equation}
\av[d] = \frac{J}{3},
\label{eq:av_d_linear}
\end{equation}

\noindent
namely, 
$\av[d]$
scales linearly with 
$J$.
The clustering coefficient for this network is zero.
Thus, from the topological point of view, 
the linear network cannot be 
considered a small world.
The rate equation for the linear metabolic network is

\begin{equation}
\frac{dn}{dt}  = g - w n(t) - 2 a n^2(t),
\label{eq:rate_linear}
\end{equation}

\noindent
leading to the steady state solution

\begin{equation}
n = {\frac{-w + \sqrt{w^2 + 8ag}}{4a}}.
\label{eq:ss_linear}
\end{equation}

\noindent
The first order correlation matrix takes the form

\begin{equation}
C = 
\left(
\begin{array}{cccccc}
0      & q     & 0      & \dots  &       & 0      \\
q      & 0     & q      &        &       & 0      \\
0      & q     & 0      & q      &       & 0      \\
\vdots &       &        & \ddots &       & \vdots \\
0      & 0     & \dots  & q      & 0     & q      \\
0      & 0     & \dots  &        & q     & 0      \\
\end{array}
\right),
\end{equation}

\noindent
where
$q = -ag/(w + 2an)^2$
[Eq. (\ref{eq:Cij})].
Using 
Eq. (\ref{eq:ss_linear}), 
one obtains

\begin{equation}
q = - { \frac{4ag}{(w + \sqrt{w^2 + 8ag})^2}}.
\end{equation}

\noindent
Since the parameters $a$, $g$ and $w$ are positive,
it is easy to see that $q$ takes values only in
the range 
$-1/2 < q < 0$.
This fact will be used in the analysis below.

To obtain the complete correlation matrix, one has to solve
Eq. (\ref{eq:Gij}). 
In the case 
of a linear metabolic network
it takes the form

\begin{eqnarray}
\left\{
\begin{array}{lrr}
\mat[G]ii = 1 & &
\\
\mat[G]ij = q(\mat[G]{i+1}{,j} + \mat[G]{i-1}{,j}) & \,\, & (i \ne j).
\end{array}
\right.
\label{eq:Gij_linear}
\end{eqnarray}

\noindent
Based on the symmetry of the problem, 
it is clear that 
for a given choice of the parameters,
the correlation between the species 
$X_i$ and $X_j$
depends only on the distance 
$d = |j - i|$
between them.
Using this indexation, 
Eq. (\ref{eq:Gij_linear}) 
becomes

\begin{eqnarray}
\left\{
\begin{array}{lrr}
G_0 = 1 & &
\\
G_d = q[G_{d+1} + G_{d-1}] &\,\, & (d \ge 1),
\end{array}
\right.
\label{eq:Gd_linear}
\end{eqnarray}

\noindent
where 
$G_d$
is the correlation matrix term for pairs of species 
$X_i$ and $X_j$
where 
$|j - i| = d$.  
Since the correlation is expected to decay exponentially 
as a function of the distance between the nodes,
we search for a solution of the form 
$G_d = \exp(-kd)$.
Inserting this expression into 
Eq. (\ref{eq:Gd_linear}) 
we obtain two possible solutions of the form

\begin{equation}
k = \ln(x \pm \sqrt{x^2 - 1}),
\label{eq:d0_linear}
\end{equation}

\noindent
where 
$x = 1/2q$.
Since the parameter $q$ is limited to the
range 
$-1/2 < q < 0$,
the parameter $x$ can take values only in the
range
$- \infty < x < -1$.
The physically relevant solution must 
satisfy the condition that the correlation 
between very distant species will vanish. 
This constraint requires that 
$|x \pm \sqrt{x^2 - 1}| > 1$.
To satisfy this condition for 
$-1/2 < q < 0$,
one has to choose the solution where the square root is subtracted.
The result is  

\begin{equation}
k = 
{\ln{
\left| 
x - \sqrt{x^2 - 1} 
\right|
}} + i\pi,
\label{eq:k_linear22}
\end{equation}

\noindent
where
$i = \sqrt{-1}$.
The correlation between species as a function of the distance 
$d$ 
between them is thus

\begin{equation}
G_d = (-1)^d 
\exp\left[
\left(
{\ln{
\left| 
x - \sqrt{x^2 - 1} 
\right|
}} 
\right)
d \right].
\end{equation}

\noindent
The pre-factor of the exponent accounts for the fact that since 
$q < 0$, 
the correlations between directly interacting species are negative.
Thus, pairs of species which are next-nearest neighbors 
in the network tend to
have positive correlations between them.
The correlation function 
[Eq. (\ref{eq:g_of_d})] 
is the absolute value of 
$G_d$, 
which comes to be

\begin{equation}
\F(d) = e^{-\frac{d}{d_0}},
\label{eq:Fcor_linear}
\end{equation} 

\noindent
where

\begin{equation}
d_0 = 
\left(
{\ln{
\left| 
x - \sqrt{x^2 - 1} 
\right|
}}
\right)^{-1}
\label{eq:k_linear2}
\end{equation}

\noindent
is the correlation length of the network.
It is interesting to examine the limit in which
$q \rightarrow 0$.
In this limit the correlations are weak and the 
typical correlation length converges to
$d_0 = -1/\ln{|q|}$.
The correlation function approaches
$\F(d) \simeq \left| q \right|^d$.
In this limit, the correlation between a pair of
species is dominated by 
the shortest path between them. 
For each step along that path, 
the correlation is multiplied by a factor of
$q$. 
Thus, the magnitude of the correlation between a pair of species
at distance $d$ from each other is approximated by 
$\left| q \right|^d$.

\begin{figure}
\includegraphics[width=12cm]{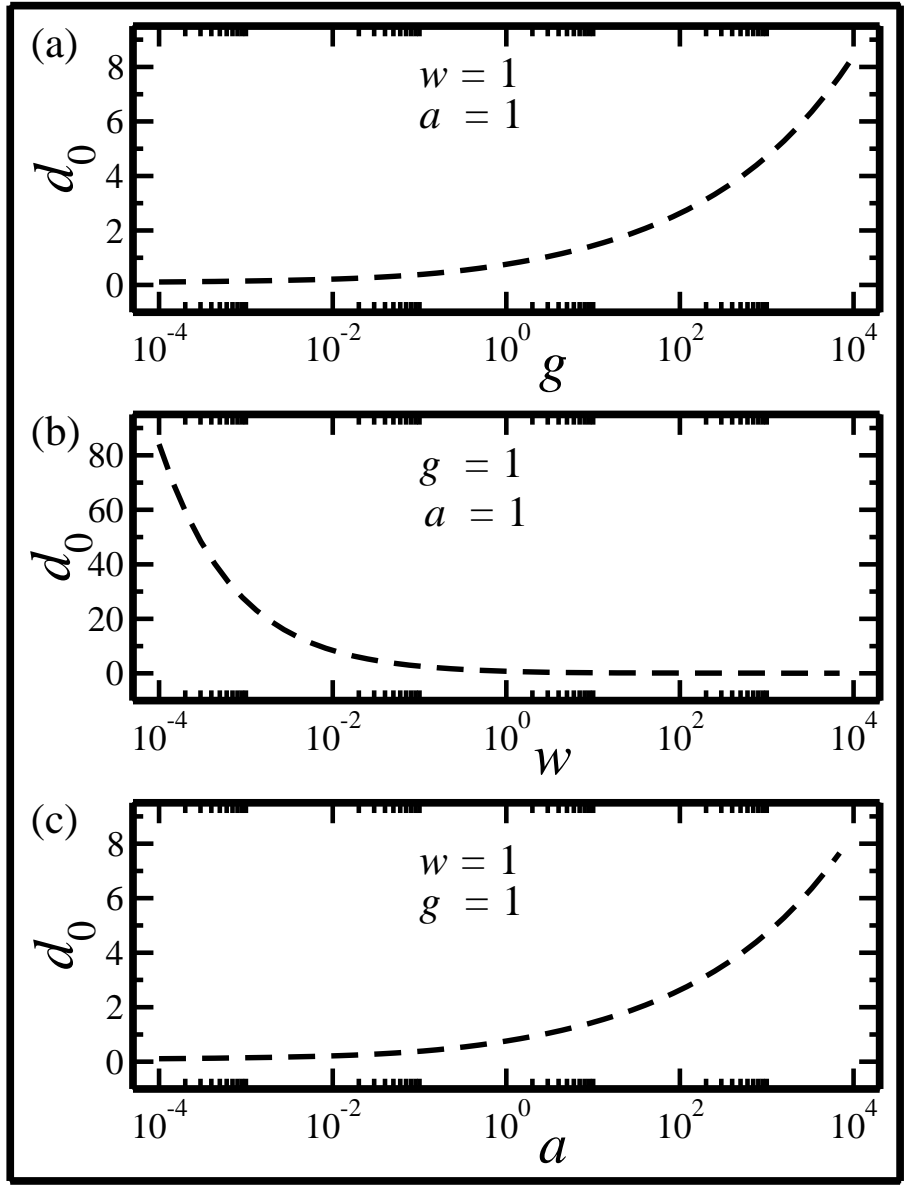}
\caption{
The correlation length,
$d_0$,
of the linear metabolic network versus
the generation rate,
$g$ (a);
the degradation rate,
$w$ (b);
and the reaction rate,
$a$ (c).
High connectivity is reached when the primary process is the reaction
process (proportional to $g$ and $a$).
The correlation length 
$d_0$
decreases with increasing 
$w$ (as the degradation becomes dominant).
}
\label{fig2_do_metabolic_graph}
\end{figure}

One can identify two limits.
In the limit where
$ag \gg w^2$ 
the correlations are strong, 
$q \rightarrow - 1/2$
and 
$d_0 \rightarrow \infty$.
In this limit, the reaction process is dominant
and long range correlations are observed.
In the limit where 
$ag \ll w^2$, 
the correlations are weak, 
and $q \rightarrow 0$.
In this limit the degradation process is dominant and
the correlation length is small.
In Fig. \ref{fig2_do_metabolic_graph} 
we present the correlation length,
$d_0$,
as a function of the parameters 
$a$, 
$w$ 
and 
$g$ 
for a linear metabolic network.
The correlation length increases with 
$a$ and $g$
(as the reaction process becomes dominant), 
and decreases with $w$
(as the process of degradation becomes dominant).

Using Eqs. 
(\ref{eq:eta}) 
and
(\ref{eq:av_d_linear}), 
the connectivity $\eta$ 
can be expressed by
$\eta = 3 d_0/J$.
The linear network clearly demonstrates the difference between the 
concepts of TSW networks and FSW networks. 
In the topological sense it is as far as a network
can be from a small world network, as the distance scales linearly 
with the network size, and the clustering coefficient is zero.
However, in the functional sense the linear network can be
a small world network, when the reaction terms are sufficiently
dominant, enabling $d_0$ to become larger than $J$.

\begin{figure}
\includegraphics[width=12cm]{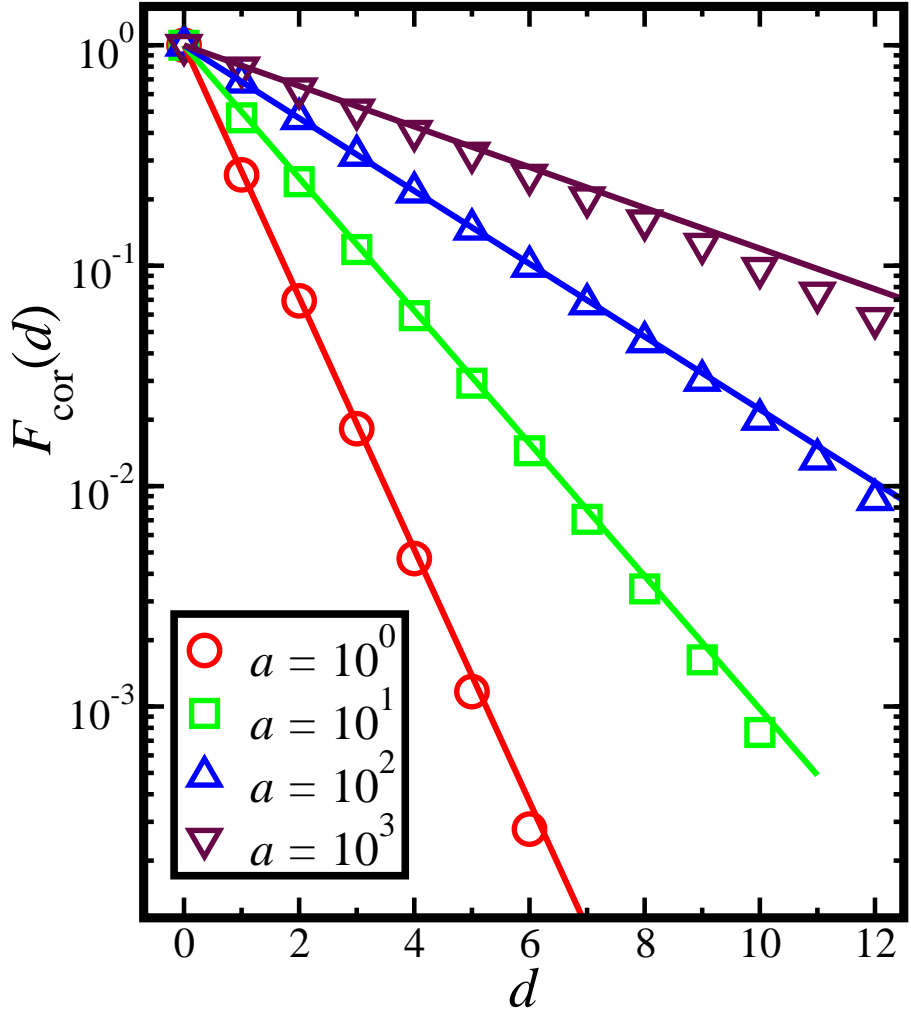}
\caption{
(color online)
The correlation function
$\F (d)$ 
for the linear metabolic network
as obtained from a numerical simulation for 
different values of the interaction rate
$a$
(symbols).
To conduct the numerical test we integrate the equations for the 
linear network and bring them to the steady state condition.
Then we force a small perturbation 
$\Delta n_0$ 
on the concentration
$n_0$ 
of the species
$X_0$.
We evaluate the correlation function using
$\F (d) = 
\left| 
\Delta n_d / \Delta n_0 
\right|$.
The correlations decay exponentially with the distance between species. 
The typical correlation length increases as the reaction rate 
is increased.
The results are in agreement with the theoretical results of 
Eq. (\ref{eq:Fcor_linear}) (solid lines).
Slight deviations appear due to the fact that
in numerical simulations
$\Delta n_0$ must be finite.
}
\label{fig3_perturbation}
\end{figure}

In order to examine the theoretical predictions of the method, 
we conducted a simulation of the long linear metabolic network described above.
In this simulation we constructed a linear network of 
$J = 100$
reacting species with periodic boundary conditions,
namely, 
$X_0$ 
reacts with
$X_{99}$. 
At time
$t = 0$
we assigned to each reacting species its steady state concentration
$n_i$. 
Then we forced the concentration
$n_0$
to be slightly above its steady state value,
namely 
$n_0(t) = n_0 + \Delta n_0$,
where
$\Delta n_0 \ll n_0$.
We then let the network relax to its new steady state. 
We denote the resulting 
change in the steady state concentration of the species
$X_i$
by
$\Delta n_i$.
In Fig. \ref{fig3_perturbation} we show the absolute value of
$\Delta n_i / \Delta n_0$
as a function of
$d$,
the distance of the node
$X_i$ 
from 
the perturbed node,
$X_0$.
These results, obtained from direct integration of the rate equations, 
are shown for different values of the reaction rate
$a$ (symbols).
When  
$a$ 
increases the typical correlation length becomes higher, and
the effect of the local perturbation of 
$X_0$
extends to more distant species.
The results are in good agreement with the theoretically derived correlation 
function, 
$\F (d)$ 
[Eq. (\ref{eq:Fcor_linear})]
(solid lines).
Slight deviations appear for distant species.
This is because in numerical simulations one must choose
$\Delta n_0$
to be a finite perturbation.
The resulting deviation in the rest of the species is thus affected 
by higher order terms in the Taylor expansion which are not accounted 
for by our method.
Here the generation rates and the 
degradation rates of all the species are
$g = 1$ and $w = 1$
respectively.
The network becomes an FSW network once
$d_0 \ge 17$,
which is approximately the average path length for this network.
This condition is satisfied for
$a \ge 2 \times 10^5$.


\subsection{Perfect Tree Network}
\label{example_2}

\begin{figure}
\includegraphics[width=12cm]{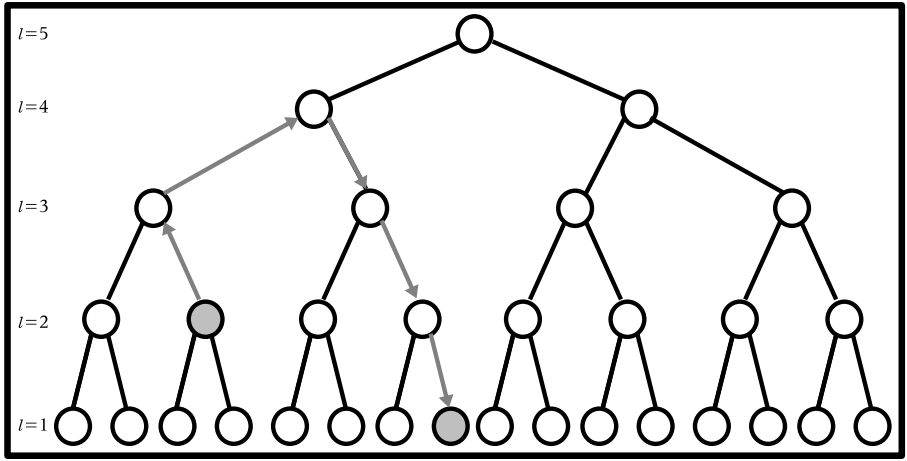}
\caption{
A tree-like network with 
$N = 5$ 
levels. 
Each node is linked with exactly one node 
at the level above it (father), and 
$m$ 
nodes at the level below (siblings).
The top node (here at level $l = 5$)
is the root node.
The order of the tree is
$m = 2$,
and the degree of the nodes is
$r = 3$.
The path between a pair of nodes is characterized 
by the number of upward steps 
followed by the number of downward steps 
to get from one node to the other.
For the path between the two shaded nodes 
$\vec d = (2,3)$.
}
\label{fig4_tree_network}
\end{figure}

Hierarchical structures are common in realistic networks.
For instance, ecological networks have 
in many cases distinct trophic levels.
Social organizations are also constructed 
in a tree-like framework.
Here we relate to a hierarchical metabolic network.
Consider a metabolic network of 
$J$
nodes where each node is assigned a level 
$l$
($l = 0, \dots, N$).
The highest level
$l = N$ 
consists of 
a single node, referred to as the {\it root}. 
Each node at level
$l$
is then connected to exactly one node at level
$l + 1$
(the {\it parent})
and 
$m$
nodes at level
$l - 1$
(the {\it siblings}).
The parameter
$m$
is defined as the order of the tree.
The degree of all the nodes 
(except those at the levels zero and $N$)
is thus
$r = m + 1$
(Fig. \ref{fig4_tree_network}).
Since this network is hierarchical, 
the up and down directions are well defined.
Stepping from a node at level 
$l$ 
to a node at level 
$l + 1$
will be considered going up the network, 
while stepping from level 
$l$
to level
$l - 1$ 
is going down the network.
Note that in a tree-like network it is not possible
to go sideways, 
as there is no edge connecting 
two nodes at the same level.
Consider a species 
$X_i$, 
which is at a distance  
$d$
from some other species 
$X_j$. 
The path between them consists of
$u$
steps up the network and 
$v$
steps down the network.
The total distance satisfies 
$d = u + v$,
and the path between them can be noted by
$\vec d = (u,v)$.
For example, the path between the two shaded nodes in 
Fig. \ref{fig4_tree_network} is
$\vec d = (2,3)$ and the distance is $d = 5$. 
Two species are said to be located in the 
same branch if in the path between them
either
$u = 0$
or
$v = 0$.
The reaction rate matrix and the first order 
correlation matrix have non-zero values only 
for directly interacting species, 
namely, for pairs of species where either
$u = 1$ and $v = 0$,
or 
$v = 1$ and $u = 0$.

In order to avoid the complexities 
related to the boundaries of the network,
we consider the case in which
$N \gg 1$.
For simplicity, 
we take the generation and the degradation rates to be
$g_i = 1$
and
$d_i = 1$
for 
$i = 1,\dots, J$.
The reaction rate is
$a_{ij} = a$
for each pair of nodes 
$i$ and $j$ 
that react with each other. 
Under these conditions, 
the network is symmetrical and the rate equations are 
identical for all nodes:

\begin{equation}
\frac{dn}{dt} = 1 - n(t) - r a n^2(t).
\label{eq:rate_tree}
\end{equation}

\noindent 
The steady state solution is thus

\begin{equation}
n = \frac{-1 + \sqrt{1 + 4 r a}}{2 r a},
\label{eq:tree_ss}
\end{equation}

\noindent 
and the non-zero elements in the first order correlation matrix  
[Eqs. (\ref{eq:2ni_2nj}) and (\ref{eq:Cij})] are

\begin{equation}
q = - \frac{4a}
{
\left(
1 + \sqrt{1 + 4 r a}
\right)^2
}.
\label{eq:q_tree}
\end{equation}

\noindent
Two limits are observed.
In the limit of strong interactions, where 
$a \gg 1$,
the matrix elements approach
$q \simeq -1/r$.
In the limit of weak interactions, where
$a \ll 1$   
one obtains
$q \simeq -a$.
In any case the values that 
$q$
can take are limited to
$-1/r \le q \le 0$.

For an infinite perfect tree with uniform rate constants,
the correlation between all pairs of species with that same values of
$u$ and $v$
are the same.
We denote this correlation by 
$G_{u,v}$.
In each line
$i$
of the first order correlation matrix there are exactly 
$r$
non-zero terms.
One term for 
$X_i$'s
parent and
$m$
terms corresponding to 
$X_i$'s
siblings.
The correlation 
$G_{u,v}$
between two species 
$X_i$ and $X_j$
is thus carried via the the parent of the species 
$X_i$, 
for which the correlation with
$X_j$
is
$G_{u - 1,v}$,
and via the 
$m$
siblings of the species 
$X_i$, 
for which the correlation with
$X_j$
is
$G_{u + 1,v}$.   
Eq. (\ref{eq:Gij}) thus takes the form

\begin{eqnarray}
\left\{
\begin{array}{lrcr}
G_{0,0} = 1 & &
\\
G_{0,v} = q \left[ G_{0,v + 1} +  G_{0,v - 1} + (m - 1)G_{1,v} \right] 
& & {\rm for} &  v>0  
\\
G_{u,v} = q(G_{u - 1,v} + m G_{u + 1,v}) & & {\rm for} & u > 0
\end{array}
\right.
\label{eq:Gij_tree}
\end{eqnarray}

\noindent
The first equation states that the correlation 
of every species with itself is unity.
The second equation accounts for the 
correlations between 
species at the same branch,
measuring the effect of variation in the higher level
node on a node at a lower level.
The third equation accounts for all the correlations
that are not included in the first two equations.
More specifically, it includes the correlations between 
species from different branches.
It also includes the correlations between pairs on 
the same branch, measuring the effect of variation in the
lower level node on a node at a higher level.
We seek a solution of the form 
$G_{u,v} = e^{- \vec k \cdot \vec d}$,
where 
$\vec k = (k_1,k_2)$ 
satisfies the condition that correlations 
vanish between distant species. From the 
third equation one obtains

\begin{equation}
k_1 = 
\ln{
\left[
\frac{1}{2q}
\left(
1 \pm \sqrt{1 - 4 m q^2}
\right)
\right]
},
\label{eq:k1_tree}
\end{equation}

\noindent
while from the second equation one obtains

\begin{equation}
k_2 = 
\ln{
\left\{
\frac{1}{2q}
\left[
\left(
\frac{x - (m - 1)q}{x} 
\right)
\pm
\sqrt{
\left(
\frac{x - (m - 1)q}{x}
\right)^2
-4q^2
}
\right]
\right\}
},
\label{eq:k2_tree}
\end{equation}

\noindent
where
$x = e^{k_1}$.
In order to satisfy the conditions that 
$G_{u,v}$
does not diverge for 
$u \gg 1$
while
$-1/r \le q \le 0$,
one has to choose the solution with the plus sign for 
$k_1$
in Eq. 
(\ref{eq:k1_tree}).
The same condition for 
$v \gg 1$
requires one to choose the solution with the plus sign for 
$k_2$
in Eq. 
(\ref{eq:k2_tree}).

\begin{figure}
\includegraphics[width=12cm]{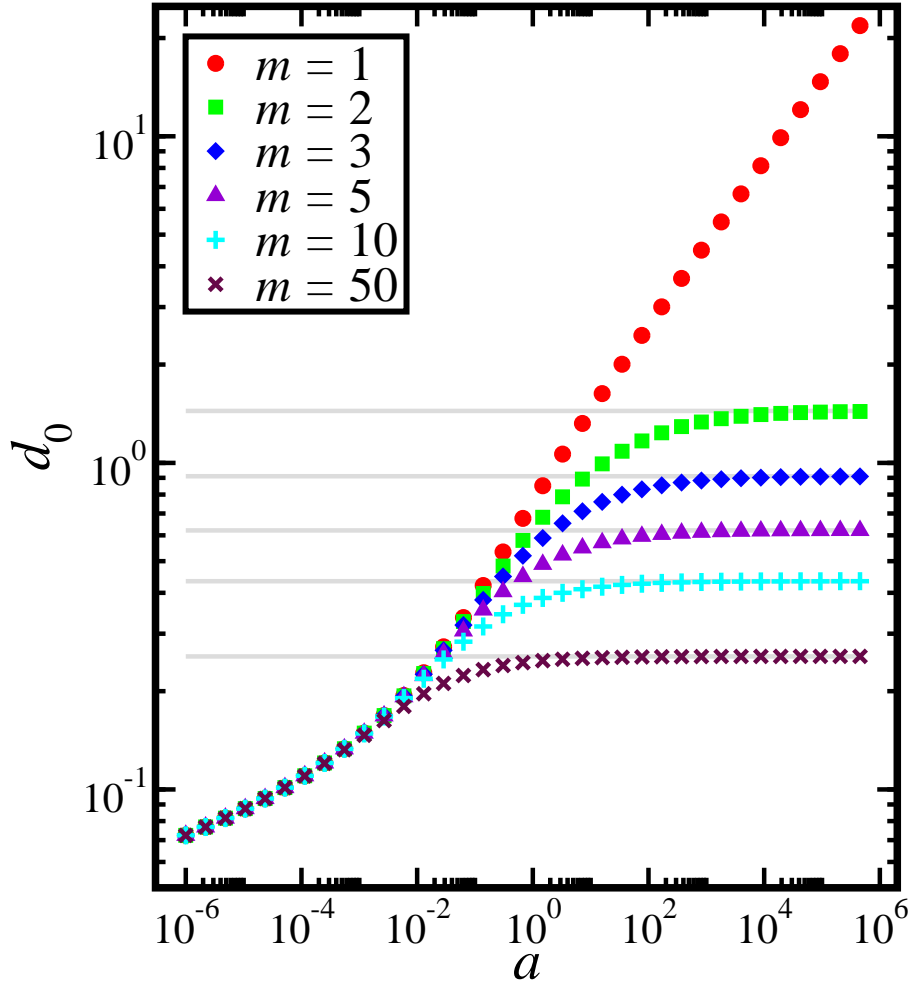}
\caption{
(color online)
The correlation length
$d_0$ 
versus the reaction rate
$a$ 
for a metabolic network with a perfect tree structure.
The results are shown for trees of different order, 
$m$
(symbols).
For 
$m = 1$
the results coincide with those obtained for the linear network.
For higher orders the correlation length is bound from above by
$d_0^{\rm max} = - 1/\ln(m)$ 
(gray horizontal lines).
Thus, for a sufficiently large tree-like network, 
where the average path length is 
larger than 
$d_0^{\rm max}$,
the connectivity is always less than unity.
Tree like networks are thus not expected to display FSW behavior.
}
\label{fig5_d0_vs_a}
\end{figure}

After some algebraic manipulations it can be shown that 
$k_1 = k_2$.
The correlation between any pair of species is thus

\begin{equation}
G_{u,v} = 
e^{i \pi d} e^{-\frac{d}{d_0}},
\label{eq:G_d_tree}
\end{equation}

\noindent
where
$d = u + v$
is the distance between the two species, and

\begin{equation}
d_0 = 
\left(
\ln{
\left|
\frac{1 + \sqrt{1 - 4 m q^2}}{2q}
\right|
}
\right)^{-1}
\label{eq:d0_tree}
\end{equation}

\noindent
is the correlation length of the tree-like network.
The correlation function is
$\F(d) = \exp(-d/d_0)$.
Note that for 
$r = 2$ ($m = 1$)
this solution coincides with the solution obtained for the linear network 
[Eq. (\ref{eq:k_linear2})].
In the limit of weak interactions, where 
$a \ll 1$
and 
$q \rightarrow 0$
the correlation function approaches
$\F(d) \simeq |q|^d$.
In this limit, due to the weak interactions,
the correlation between a pair of species is dominated by
the shortest path between them. 
In the limit of strong interactions, where
$a \gg 1$
and
$q \rightarrow -1/r$,
the correlation length satisfies
$d_0 \rightarrow 1/ \ln (m)$.
For 
$m > 1$
the correlation length is always finite. 
Since the average path length of 
a perfect tree-like network must scale in some form 
with the number of levels in the tree, 
one obtains that for a large enough tree network
the connectivity will always be less than unity.
Thus a perfect tree-like network of order 
$m=2$ 
or more will never be an FSW.
In Fig. \ref{fig5_d0_vs_a} we show the correlation length
$d_0$
as obtained for a metabolic network with a perfect tree topology
vs. the reaction rate 
$a$ 
(symbols).
The results are shown for trees of different orders.
Here
$g = d = 1$,
and 
$a$
is varied.


\section{More Complex Networks}
\label{complex_networks}

To demonstrate the applicability of the NCF method,
we now refer to the analysis of a set of more complex networks.
Here analytical solutions are not available, and the correlation matrix 
must be obtained numerically.
We analyze three different topologies following the structural classification
proposed by Estrada 
\cite{Estrada2007}.
The first example represents a class of networks which are organized into 
highly connected modules with few connections between them.
The second example will be of a network with a highly connected central core
surrounded by a sparser periphery,
and the last example will be of a scale-free network.

\begin{figure}
\includegraphics[width=12cm]{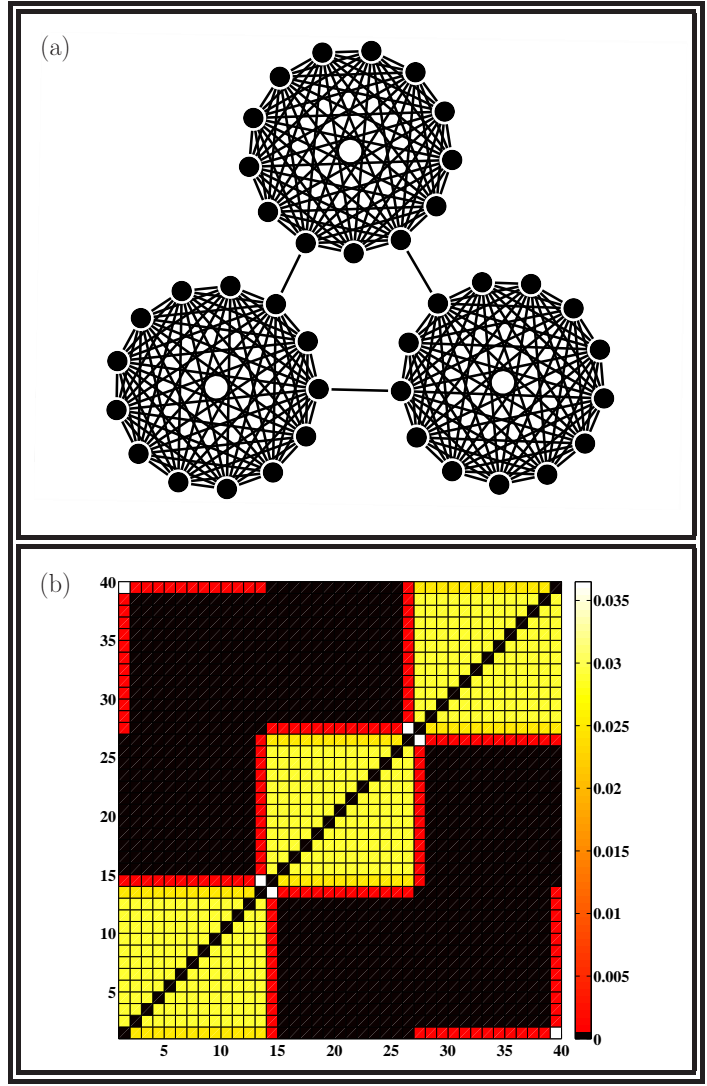}
\caption{
(color online)
(a) A network constructed of three fully connected modules,
with single bonds between them.
(b) The correlation matrix features high correlations within 
the modules, and very small correlations between pairs of nodes 
from different modules.
The matrix is constructed of three almost uncoupled blocks,
reflecting the near bipartite topology of the network.
The diagonal terms, which are all unity, do not appear in the Figure.
}
\label{fig6_three_joint_clusters}
\end{figure}

Consider a network constructed of three fully connected modules (communities),
with a single connection between each pair of communities.
This network is displayed in Fig. \ref{fig6_three_joint_clusters}(a).
Here, each community consists of 
$13$ 
nodes, adding up to a total of
$J = 39$
nodes.
To obtain,
$n_i$, $i=1,\dots,39$,
the steady state solution for the concentrations of the different 
reacting species we solve 
Eq. (\ref{eq:metabolic_rate}) using a standard Runge-Kutta stepper.
The parameters we use are
$g_i = 1$
and
$d_i = 1$.
The reaction rate
$a$
between pairs of reacting species is also set to unity.
We then construct the first order correlation matrix,
$\mat[C]ij$,
as appears in Eq. (\ref{eq:Cij}).
The complete correlation matrix, 
$\mat[G]ij$, 
is obtained from
Eq. (\ref{eq:Gij}). 
It
consists of a set of 
$39 \times 39$
linear algebraic equations.
Solving these equations,
one obtains the complete correlation 
matrix of the network.
For this network, the main insight on the global functions of the 
network can be deduced from the complete correlation matrix,
which is displayed in Fig. \ref{fig6_three_joint_clusters}(b).
The diagonal terms, which are all unity,
are omitted from the Figure.
As expected, strong correlations appear between species within the
same community (sub-matrices along the diagonal),
and vanishingly small correlations appear between species from 
different communities.
In fact, the correlation matrix is close to be a partitioned block
matrix, except for a few coupling terms between the blocks.
In this case the correlation matrix reflects the topological structure
of the network, which is almost fully partitioned into 
three isolated communities.

\begin{figure}
\includegraphics[width=12cm]{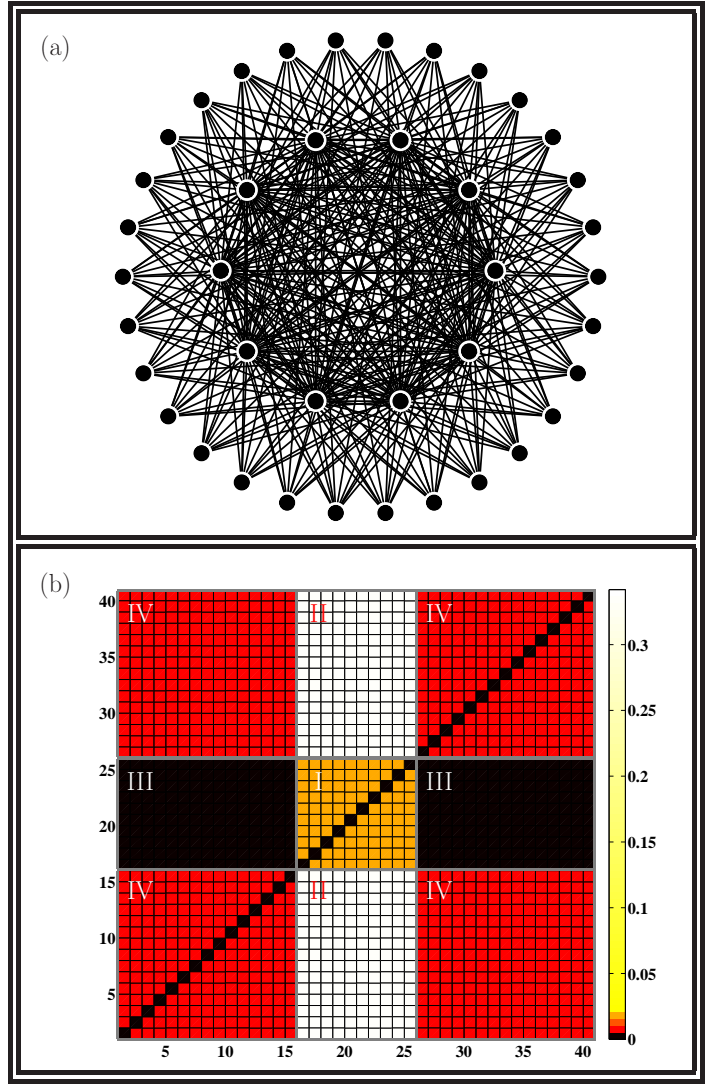}
\caption{
(color online)
(a) A network consisting of a dense fully connected core
with a sparse periphery.
The peripheral nodes are each connected to all the central ones,
but not to each other.
(b) The correlation matrix shows strong correlations between pairs
of species from the core (domain I).
The strongest dependence is between nodes from the periphery
to nodes from the core (domains II).
However nodes from the core are almost not affected by
nodes from the periphery (domains III).
Interestingly, the correlations between pairs of nodes from the 
periphery are not so low, even though they are not directly 
connected to one another (domains IV).
The diagonal terms, which are all unity, do not appear in the Figure.
}
\label{fig7_core_and_periphary}
\end{figure}

We now consider a network, 
which features a highly connected central core
surrounded by a sparser periphery. 
This network consists of
$J = 40$
nodes.
The nodes 
$X_{i}$, $i=16,\dots,25$,
are a fully connected cluster (the core),
while the 
$30$
additional nodes are connected to all the nodes in the core,
but not to each other (the periphery).
This network is shown in Fig. \ref{fig7_core_and_periphary}(a).
Following the same procedure described above one obtains 
the correlation matrix for this network 
[Fig. \ref{fig7_core_and_periphary}(b)].
The central square (domain I) shows the correlations between
the nodes in the central core.
Domains II show the correlations between peripheral nodes 
and central ones. 
The value of these correlations is high, expressing the strong
dependence of the peripheral nodes on the nodes in the central core.
On the other hand for the correlations between the central nodes 
and the peripheral ones (domains III) one obtains very low correlations.
This is an expected result, as deviations in the population of a node
from the periphery should have almost no effect on a node from the core.
An interesting result appears in domains IV.
These domains show the correlations between pairs of nodes that are both 
from the periphery.
It turns out that the effect of these nodes on each other is
stronger than the effect they have on their adjacent nodes from the
core.
This is even though the topological distance between peripheral nodes is 
$d = 2$,
while the distance between them and the central nodes is
$d = 1$.
A small perturbation in a peripheral node results in a very minor    
effect on all the central nodes.
However this minor change in the core results in a more dramatic
effect on all the rest of the peripheral nodes.
This non-trivial result exemplifies the importance of the functional methodology
as a complimentary analysis to the common topological approach.      
In the two examples shown above, we focused on the insights 
provided by the complete correlation matrix.
Below we show an additional numerical example, where
we continue the analysis to obtain the correlation length,
$d_0$,
and the connectivity
$\eta$.

One of the common 
characteristics 
of many realistic networks is 
their degree distribution that follows a power law, namely
$P(k) = \alpha k^{-\lambda}$, 
where
$\alpha$ and $\lambda$ 
are positive constants
\cite{Barabasi1999,Albert2000}.
Ecological networks, social networks and metabolic networks are
characterized by power-law degree distributions,
and are referred to as scale-free networks.
Such networks include some nodes, called hubs, 
with a degree which is orders of magnitude higher than the
average degree in the network.
Scale free networks are considered as 
highly connected, because due to these hubs 
the average path length 
between nodes is small.
In fact, in metabolic networks the average path 
length was found to be as small as
$\av[d] \simeq 3$
\cite{Jeong2000}.
Below we examine a scale free network which is a TSW network, and
determine whether it is also an FSW network.

\begin{figure}
\includegraphics[width=12cm,angle=-90]{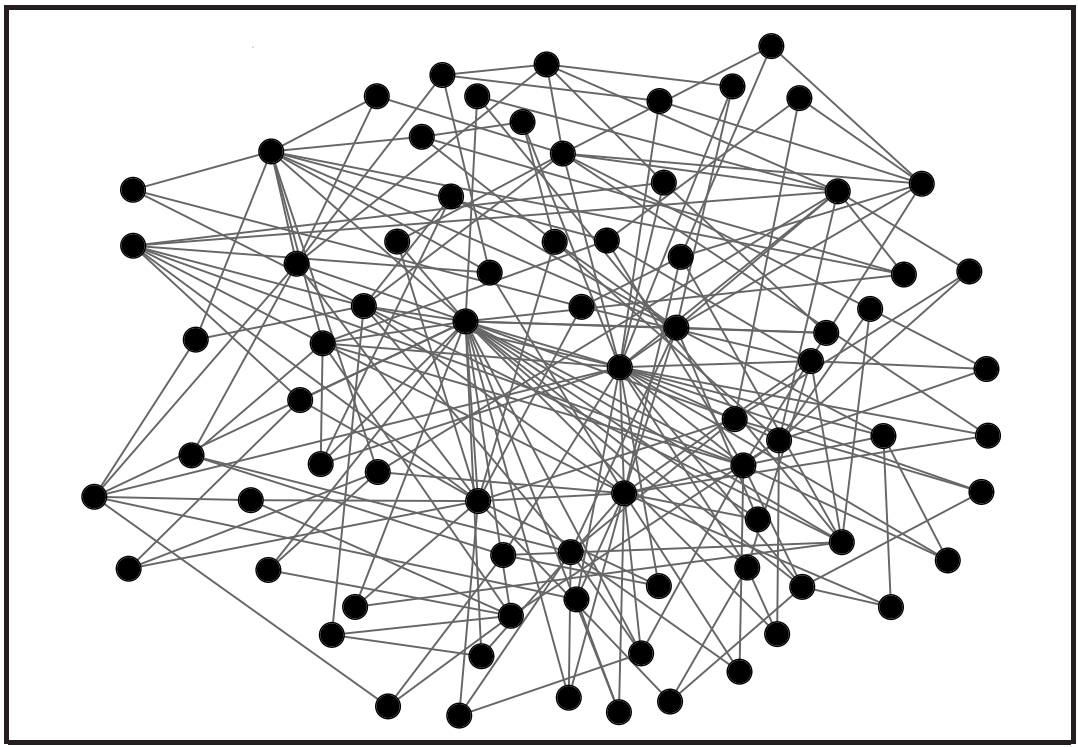}
\caption{
Scale free network
consisting of 75 reacting species,
constructed using the preferential attachment algorithm.
The average path length of this network is
$\av[d] = 2.43$ 
and its diameter is 
$D = 4$.
}
\label{fig8_scale_free_network}
\end{figure}

To construct a scale free network we use the preferential attachment algorithm
\cite{Barabasi1999}.
In this algorithm a single new node is added at each iteration and 
$m$
edges are drawn from it to the set of existing nodes.
The probability of linking the new node to some existing node
$X_i$
is proportional to the current degree of the node 
$X_i$.
This way, nodes which already have a higher degree than others
have a high probability of obtaining more links and becoming hubs.
Here we constructed a scale free network consisting of
$J = 75$
nodes.
The number of edges added in each iteration was 
$m = 3$.
The result is the graph appearing in Fig. \ref{fig8_scale_free_network}.
The diameter of this network is
$D = 4$
and its average path length is
$\av[d] = 2.43$.

\begin{figure}
\includegraphics[width=12cm]{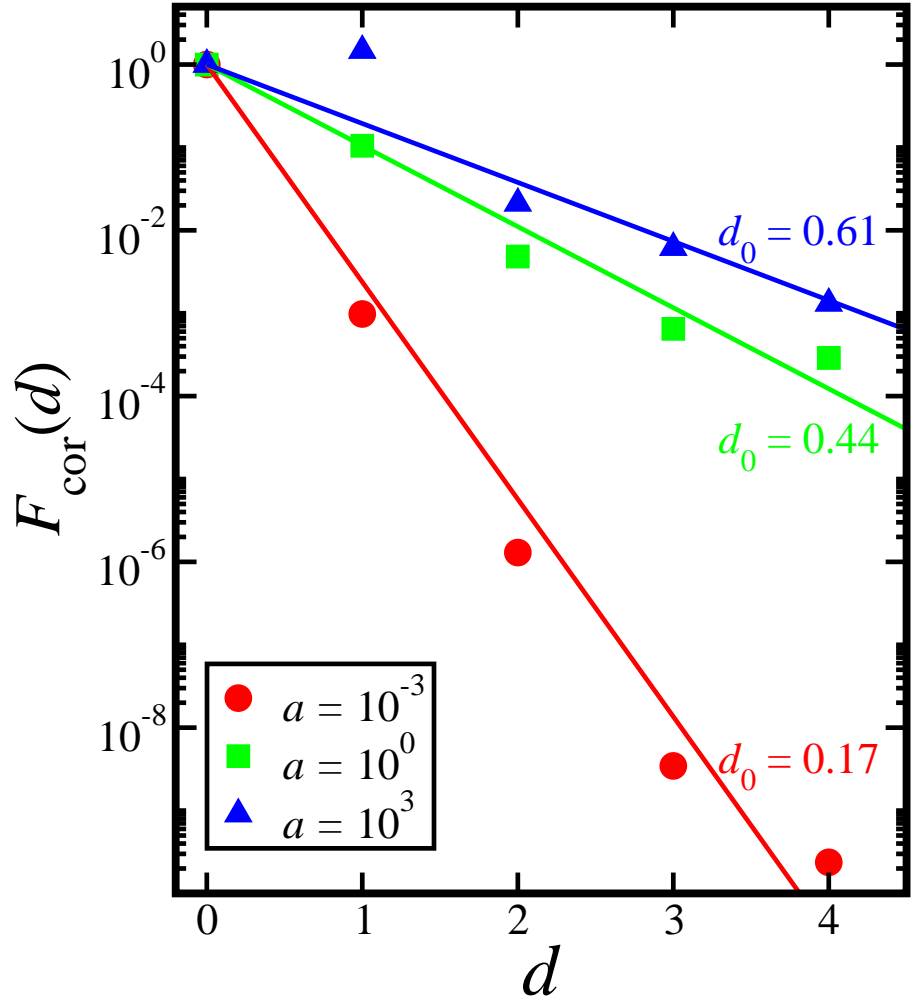}
\caption{
(color online)
The correlation function vs. distance as obtained for the 
scale free network appearing in 
Fig. \ref{fig8_scale_free_network}
for different values of the parameter
$a$ (symbols).
The correlations decay rapidly for low values of 
$a$,
and more gradually for large values of 
$a$.
By fitting the correlation function to an exponential (solid lines),
the correlation length
$d_0$ can be obtained.
}
\label{fig9_g_of_d_scale_free}
\end{figure}

\begin{figure}
\includegraphics[width=12cm]{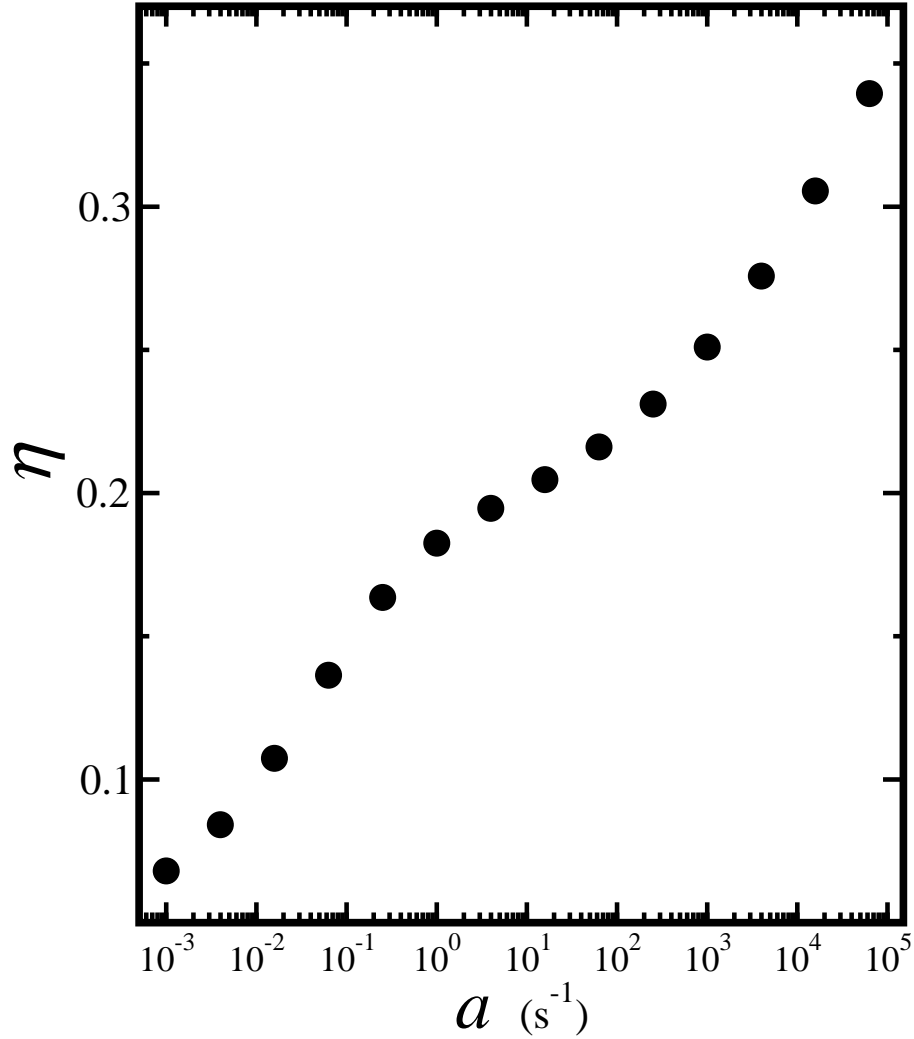}
\caption{
The connectivity,
$\eta$
vs. 
$a$
as obtained for the scale free network shown in 
Fig. \ref{fig8_scale_free_network}. 
The connectivity increases logarithmically as a function of
$a$. 
Note that 
$\eta < 1$
for a very broad range of values of the parameter $a$.
This implies that although scale-free networks are commonly TSW networks,
in the functional sense they may not be FSW networks.  
}
\label{fig10_eta_vs_a_scale_free}
\end{figure}

Solving Eq. (\ref{eq:metabolic_rate}) 
we obtain the steady state solution for the concentrations of all the 
reactive species.
The parameters are
$g_i = 1$
and 
$d_i = 1$
for 
$i = 1,\dots,75$.
The reaction rate
$a$
between pairs of reacting species is varied.
In this case, obtaining the complete correlation matrix, 
$\mat[G]ij$, 
requires the solution of
$75 \times 75$
linear algebraic equations
[Eq. (\ref{eq:Gij})].
We solve these equations and then average 
over the correlations between equidistant species
to obtain the correlation function 
$F_{\rm cor}(d)$
[Eq. (\ref{eq:g_of_d})].
In Fig. \ref{fig9_g_of_d_scale_free}
we show the resulting correlation function
$F_{\rm cor}(d)$
vs.
$d$
for three different values of the reaction rate
$a$ (symbols).
When the interaction is suppressed 
($a \ll 1$) 
the correlations decay rapidly. 
When the interaction is dominant 
($a \gg 1$), 
correlations are maintained over 
long distances.
By fitting the correlation functions to exponential functions (solid lines) 
one obtains the typical correlation length,
$d_0$,
and the connectivity,
$\eta$,
of each of the networks.
The results for 
$\eta$
vs. the reaction rate
$a$ 
are shown in Fig. \ref{fig10_eta_vs_a_scale_free}. 
It is found that the connectivity increases logarithmically 
with 
$a$.
Note that for a very wide range of 
values of the parameter $a$,
the connectivity remains lower than unity.
This means that although the examined scale free network 
is a TSW, for a very wide 
range of parameters it is not an FSW.
Only in the extreme cases of very strong interactions 
FSW behavior might emerge.


\section{Summary and Discussion}
\label{sec_summary}

We have presented the NCF method for the analysis and evaluation of  
the connectivity of interaction networks. 
The method complements the topological analysis of networks, 
taking into account the functional nature of the interactions 
and their strengths. 
The method enables to obtain the
correlation matrix, which provides the correlations between 
pairs of directly and indirectly interacting nodes. 
In certain cases, one may gain insights on the network's 
functionality by writing down the complete correlation matrix.
For instance, one can identify domains of high and low correlations.
In other cases it is more insightful to extract
the macroscopic characteristics of the network
from the matrix.
In particular, we have shown how to calculate the 
typical correlation length of the network.
This correlation length, which has to do with the functionality
of the network, can be compared to topological characteristics such
as the average minimum path length of the network.
The ratio between these two lengths
provides the characteristic connectivity of the network.
It was shown that the topological analysis alone is not sufficient
in order to characterize the 
functionality of a network.
For instance, networks with small world 
topology may display low connectivity, while 
networks that do not exhibit
small world topology may display high connectivity.
This is because in terms of the functionality of the network, 
when the correlation length is large, 
even distant species may be highly correlated. 
We demonstrated the method for metabolic networks with different 
topological structures, and
identified the regimes of low connectivity and of high connectivity.
As expected, these regimes depend on topological features, 
such as the number of 
species or the average minimum path length between pairs of species.
However, they also depend 
on functional features such as the type of interactions in the network 
and the rate constants of the different processes.

The NCF method was demonstrated for metabolic networks, but its 
applicability is much wider.
In fact, the method could be applied to any reaction network
that can be modeled by rate equations.
Such networks include 
metabolic networks
\cite{Peacocke1989},
chemical networks
\cite{Tielens2005,Gillespie2007},
gene expression networks
\cite{Palsson2006,Alon2006}
and ecological networks
\cite{Murray1989}.
It is common to use rate equations for the modeling of these
types of networks.
In certain models of social networks,
the flow of information as well as the spreading of viruses
can also be described by rate equations.
The method is not suitable for obtaining the correlations 
in Ising type models, where the nodes are assigned discrete variables,
which cannot be modeled using continuous equations.
The number of elements in the correlation matrix is equal to the 
number of pairs of nodes in the system.
When applying the NCF method, 
one writes a single linear equation
for each matrix element.
Thus, from a computational point of view, the scaling of the NCF method is
quadratic in the number of reactive species.
This enables the application of the method to networks which include
even thousands of nodes.
It is straightforward to extend the application of the method 
to the other types of interaction networks mentioned above.
A few examples are addressed below.

Consider, for example, gene expression networks. 
These networks consist of genes and proteins that interact with each other.
In addition to protein-protein interactions, 
already analyzed in the 
context of metabolic networks, genetic networks include
transcriptional regulation processes,
where some genes regulate
the expression of other genes.
In recent years,
much information has been acquired about the topology
of these networks,
for certain organisms 
such as {\it Escherichia coli}
\cite{Alon2006}.
The problem is that these networks are very elaborate, 
and may consist of thousands of nodes. 
This limits our ability to simulate their 
functionality, and thus, currently
most of the theoretical and computational 
analysis of these networks is focused
on small modules
\cite{Alon2006}.
In this analysis, one performs 
simulations of small subnetworks consisting of only a few nodes.
These subnetworks are expected to play specific 
roles in the functionality of the 
network as a whole.
Such approach is valid if
an isolated module
maintains its function when
incorporated in a large network in which it interacts with 
many other genes.
We expect the analysis presented here to 
provide some insight on this matter.
By obtaining the complete correlation matrix, one 
can characterize the dependence of different
proteins and genes on one another.
The network may then be divided into subnetworks, 
grouping together nodes that are highly 
correlated, and excluding ones that are not.
It is expected that
these modules will not function significantly 
differently when analyzed in the context of the surrounding
network nodes.
In addition, the typical correlation length 
will provide us with an approximate radius 
beyond which correlations may be neglected.
To simulate a module properly, 
one needs to include all the nodes which are
within that radius from the module.
Other possible applications regard social networks.
For instance, the process of viral spreading could be analyzed
\cite{PastorSatorras2001,Lind2007}.
Many social networks are known to be small world networks. 
However, this does not mean that any contagious disease 
spreads rapidly.
This is, possibly, because for certain diseases 
the correlation length is small.
Using the method presented here, 
one can obtain this correlation length, taking into account
the specific rate constants of the viral flow.

The recent applications of graph theory to many natural macroscopic systems
was enabled by focusing on their topology.
This approach has been very fruitful, 
as it uncovered the mutual structure of networks from
many different fields.
In particular, the ubiquity of the scale 
free degree distribution, and the small world
topology was found.  
However, it still is not completely clear what functional 
meaning can be given to these 
topological properties in different contexts. 
A recently proposed approach derives the key 
aspects of the network functionality
from its topological structure
\cite{Stelling2002}.
Other approaches use the Ising Hamiltonian to 
describe the interaction pattern
between nodes on scale free and small world networks
\cite{Jeong2003,Pekalski2001}.
Functional characteristics such as phase transitions, 
and critical exponents are 
then observed.
The NCF method presented in this paper complements these approaches.
It can be applied to a variety of different interaction processes,
such as metabolic, ecological or social interactions, all of which
can be described by rate equations.
We believe that the approach presented here 
will lead to new insights on the behavior of networks
and their functionality.


\clearpage
\newpage


\clearpage
\newpage
 
%
%

%
%

%
%

%
%

%
%

%
%

%
%

%
%

%
%

%
%

\end{document}